\documentclass[iop,apj]{emulateapj}
\usepackage{booktabs}

\newcommand{\kms}{\mbox{$\mathrm{km\,s^{-1}}$}}
\newcommand{\Ion}[2]{#1{\,\scriptsize #2}}
\newcommand{\Teff}{\mbox{$T_{\mathrm{eff}}$}}
\newcommand{\Tbb}{\mbox{$T_{\mathrm{bb}}$}}
\newcommand{\Lines}[3]{\Ion{#1}{#2}\,$\lambda\lambda$\,#3}

\begin{document}
\title{Time-Variable Aluminum Absorption in the Polar AR Ursae
  Majoris, and an updated estimate for the mass of the white dwarf}

\author{Yu Bai\altaffilmark{1}, Stephen Justham\altaffilmark{2,1}
   JiFeng Liu\altaffilmark{1,2}, JinCheng Guo\altaffilmark{1,2},
   Qing Gao\altaffilmark{1,2}, Hang Gong\altaffilmark{1,2},
   }

\altaffiltext{1}{Key Laboratory of Optical Astronomy, National Astronomical Observatories, Chinese Academy of Sciences,
       20A Datun Road, Chaoyang Distict, 100012 Beijing, China; ybai@nao.cas.cn}
\altaffiltext{2}{College of Astronomy and Space Sciences, University
  of the Chinese Academy of Sciences, Beijing, China 100049}

\begin{abstract}
We present spectra of the extreme polar AR Ursae Majoris (AR UMa) which display a clear
\Ion{Al}{I} absorption doublet, alongside spectra taken less than
a year earlier in which that feature is not present.
Re-examination of earlier SDSS spectra indicates that the
\Ion{Al}{I} absorption doublet was also present $\approx$8 years
before our first non-detection.
We conclude that this absorbing material is unlikely to be on
  the surface of either the white dwarf (WD) or the donor
  star. We suggest that this \Ion{Al}{I} absorption feature arises in
  circumstellar material, perhaps produced by the evaporation of
  asteroids as they approach the hot WD.  The presence of any
  remaining reservoir of rocky material in AR UMa might
  help to constrain the prior evolution of this unusual binary system.
We also apply spectral decomposition to find the stellar parameters of
the M dwarf companion, and attempt to dynamically measure the mass of
the WD in AR UMa by considering both the radial velocity curves of
the H$_\beta$ emission line and the \Ion{Na}{I} absorption line.
Thereby we infer a mass range for the WD in AR UMa of
0.91 $M_{\odot}$ $<$ $M_{\mathrm{WD}}$ $<$ 1.24 $M_{\odot}$.
\end{abstract}


\section{Introduction}
The polar AR UMa is inferred to contain a white dwarf (WD) that possesses a magnetic field of
$\sim$240 MG -- the strongest of the known polars
\citep[][]{Schmidt96,Gansicke01}.
Its optical observations show that it often changes between high and low brightness states,
remaining in the low state for approximately 90\% of the time. During the low
state, ellipsoidal variations reveal an orbital period
of 1.932 hr \citep{Remillard94}, and both the WD and the M dwarf (MD)
donor star can be detected from optical spectra.

Here we report the detection of a time-variable aluminum absorption
line in spectra of AR UMa.  We conclude that the line probably originates from
circumstellar material around the WD, and tentatively suggest that the most likely
origin of this material is from evaporation of infalling asteroids.

Similar scenarios have been examined in recent
  years, with both observational evidence and theoretical studies
  finding that tidal disruption and evaporation of rocky asteroids
  can account for circumstellar material around WDs
  \citep{Lallement11,Debes12,Perez13,Veras14,Veras15,Vanderburg15}.
This suggests that
  reservoirs of asteroids, comets, and planets/planetisimals may exist
  in wide orbits around some WDs. If so, their orbits could
  occasionally become sufficiently unstable or elliptical that they
  are disrupted as they approach the WD
  \citep{Frewen14,Stone15,Veras15}.  Rocky debris may also be
  evaporated by the hot radiation from a WD into circumstellar gas, from which heavy element
  absorption lines can originate
  \citep{Lallement11,Debes12,Dickinson12,Perez13}. Such scenarios
  may also explain the features of non-degenerate stars -- notably $\beta$ Pictoris, a star with a
  well-established debris disk
  \citep[see, e.g.,][]{Beust+1990,Beust+1996,Karmann+2003}.  Material from an evaporating rocky planet may
  explain photometric variations in Kepler observations of the K dwarf
  KIC 12557548 \citep{Rappaport+12}.

Our detection was serendipitous.  Despite being a very well-known and
extreme system, the basic parameters of the stellar components in AR
UMa are poorly constrained. The emission lines from the heated inner
hemisphere of the MD are a poor tracer of the barycenter of the MD,
and thus have not enabled precise dynamical mass measurements. Using
the H$_{\alpha}$ emission line, \citet{Schmidt99} estimated
the mass of the WD as 0.4 $\leq$ M$_{\mathrm{WD}}$ $\leq$ 1.0
M$_\odot$.  This observing campaign was initially intended to measure the system
parameters by taking advantage of the absorption lines from the MD, such as
\Ion{Na}{I}, which are more isotropically distributed over the
photosphere of the MD \citep{Rebassa07}.

Our observations of AR UMa and our updated orbital ephemeris are briefly described in Section~2,
the detection of aluminum absorption lines is presented in
Section~3, and our new constraints on the system masses and geometry are given in Section~4.
Section~5 features potential explanations for the origin of
  the aluminum absorption lines.
A summary is given in Section~6.

\begin{deluxetable*}{ccccccccc}
\tablecaption{Log of Observations \label{log}}
\tablehead{
\colhead{UT Date}&\multicolumn{2}{c}{Number of Exposures}&
\multicolumn{2}{c}{Exposure Time} & \multicolumn{2}{c}{Wavelength
  Coverage ($\lambda$$\lambda$)} & \multicolumn{2}{c}{Spectral Resolution ($\Delta$$\lambda$)}  \\
\cmidrule(l){2-3} \cmidrule(l){4-5} \cmidrule(l){6-7}  \cmidrule(l){8-9}
\colhead{} & \colhead{Blue}& \colhead{Red} & \colhead{Blue} & \colhead{Red} & \colhead{Blue} & \colhead{Red} & \colhead{Blue} & \colhead{Red} \\
\colhead{(yyyymmdd)}& \colhead{} & \colhead{} &\colhead{(s)}&\colhead{(s)}& \colhead{(\AA)} & \colhead{(\AA)} & \colhead{(\AA)} & \colhead{(\AA)} \\
}
\startdata
20120322 & 4 & 2 & 300 & 600 & 3990$-$5540 & 5680$-$9010 & 1.4 & 2.8 \\
20120518 & 2 & 2 & 600 & 600 & 3710$-$5260 & 5660$-$8900 & 1.4 & 2.8 \\
20130312 & 10& 10& 600 & 600 & 3800$-$5350 & 8390$-$9040\tablenotemark{a} & 1.4 & 1.4
\enddata
\tablenotetext{a}{Unfortunately the CCD in the red side of DBSP was
  broken; these observations used the smaller backup CCD.}
\end{deluxetable*}

\begin{deluxetable*}{rcrrcrr}
\tablecaption{Line-fitting Results \label{Pro}}
\tablehead{
\colhead{} & \multicolumn{2}{c}{Al} & \multicolumn{2}{c}{H$_{\beta}$} & \multicolumn{2}{c}{\Ion{Na}{I}} \\
\cmidrule(l){2-3} \cmidrule(l){4-5} \cmidrule(l){6-7}
\colhead{HJD$-$2456000\tablenotemark{a}}& \colhead{Velocity} & \colhead{FWHM} & \colhead{Velocity} & \colhead{FWHM} &\colhead{Velocity}&\colhead{FWHM} \\
\colhead{(day)}        & \colhead{(\kms)}     & \colhead{(\AA)} & \colhead{(\kms)}     & \colhead{(\AA)} &\colhead{(\kms)}&\colhead{(\AA)}
}
\startdata
$-$2938.561& 22.1 $\pm$ 25.2 & 1.5 $\pm$ 1.2, 2.6 $\pm$ 1.7 &                    &               &                     &                                  \\
  8.662    &                \multicolumn{2}{c}{No spectral coverage}                  &   231.0 $\pm$ 6.5  & 3.6 $\pm$ 0.3 &    280.1 $\pm$ 24.8 &  6.9 $\pm$ 1.6, 12.7 $\pm$ 2.8   \\
  8.666    &                 \multicolumn{2}{c}{No spectral coverage}                   &   175.0 $\pm$ 4.0  & 3.3 $\pm$ 0.2 &                     &              \\
  8.670    &                 \multicolumn{2}{c}{No spectral coverage}                     &   108.7 $\pm$ 2.8  & 3.2 $\pm$ 0.1 &  189.9 $\pm$ 19.0 & 16.1 $\pm$ 4.9,  7.0 $\pm$ 1.2   \\
  8.674    &                \multicolumn{2}{c}{No spectral coverage}                    &    40.7 $\pm$ 2.5  & 3.0 $\pm$ 0.1 &                     &              \\
 65.752    &                 &   $<$ 1.4\tablenotemark{b}   & $-$24.4 $\pm$ 4.1  & 3.6 $\pm$ 0.2 & $-$105.5 $\pm$ 27.8 &  7.1 $\pm$ 1.7, 17.9 $\pm$ 4.5    \\
 65.759    &                 &   $<$ 1.1\tablenotemark{b}  &$-$162.3 $\pm$ 4.8  & 3.3 $\pm$ 0.2 & $-$249.5 $\pm$ 18.9 &  5.5 $\pm$ 1.4, 21.6 $\pm$ 5.9    \\
363.650    & 70.2 $\pm$ 21.9 & 3.8 $\pm$ 1.3, 4.9 $\pm$ 1.2 &   200.4 $\pm$ 25.9 & 5.7 $\pm$ 1.1 &                    \multicolumn{2}{c}{No spectral coverage}              \\
363.683    & 77.4 $\pm$ 18.7 & 2.8 $\pm$ 0.9, 4.6 $\pm$ 1.2 &  $-$6.2 $\pm$ 4.2  & 4.0 $\pm$ 0.2 &                     \multicolumn{2}{c}{No spectral coverage}               \\
363.721    & 63.4 $\pm$ 11.3 & 4.0 $\pm$ 0.7, 3.6 $\pm$ 1.6 &    67.7 $\pm$ 37.4 & 8.6 $\pm$ 1.6 &                 \multicolumn{2}{c}{No spectral coverage}              \\
363.761    & 42.5 $\pm$ 27.5 & 4.8 $\pm$ 1.4, 5.1 $\pm$ 2.0 &    45.7 $\pm$ 3.9  & 3.9 $\pm$ 0.2 &                   \multicolumn{2}{c}{No spectral coverage}              \\
363.799    & 44.9 $\pm$ 14.8 & 3.9 $\pm$ 0.8, 4.6 $\pm$ 0.9 & $-$17.1 $\pm$ 20.1 & 3.8 $\pm$ 0.8 &                  \multicolumn{2}{c}{No spectral coverage}                \\
363.837    & 48.8 $\pm$ 17.0 & 3.8 $\pm$ 1.3, 3.8 $\pm$ 0.8 &   119.1 $\pm$ 2.8  & 4.1 $\pm$ 0.1 &                  \multicolumn{2}{c}{No spectral coverage}               \\
363.875    & 49.4 $\pm$ 13.1 & 3.3 $\pm$ 0.7, 4.9 $\pm$ 0.8 & $-$72.9 $\pm$ 17.9 & 4.5 $\pm$ 0.7 &                \multicolumn{2}{c}{No spectral coverage}              \\
363.913    & 59.8 $\pm$ 14.1 & 2.8 $\pm$ 0.7, 3.5 $\pm$ 0.8 &   200.9 $\pm$ 4.6  & 3.8 $\pm$ 0.2 &                 \multicolumn{2}{c}{No spectral coverage}               \\
363.950    & 65.5 $\pm$ 16.3 & 3.0 $\pm$ 0.8, 3.8 $\pm$ 1.0 &$-$168.7 $\pm$ 17.3 & 5.2 $\pm$ 0.7 &               \multicolumn{2}{c}{No spectral coverage}             \\
363.976    & 77.4 $\pm$ 12.5 & 4.2 $\pm$ 0.8, 3.2 $\pm$ 0.6 &   234.7 $\pm$ 10.1 & 4.1 $\pm$ 0.4 &                  \multicolumn{2}{c}{No spectral coverage}
\enddata
\tablenotetext{a}{Heliocentric Julian date.}
\tablenotetext{b}{For the strongest individual line from the doublet.}
\end{deluxetable*}

\begin{deluxetable}{llll}
\tablecaption{Fitting parameters of radial velocities \label{sin}}
\tablehead{
\colhead{Line} & \colhead{$K$} & \colhead{$\gamma$} & \colhead{P}  \\
\colhead{}     & \colhead{(\kms)}    & \colhead{(\kms)}   & \colhead{(hr)}
}
\startdata
Al           & 17 $\pm$ 9   & 61 $\pm$ 6 & 8.891 $\pm$ 0.001 \\
H$_{\beta}$  & 248 $\pm$ 4  & 34 $\pm$ 5 &
1.93201522\tablenotemark{a}\\
\Ion{Ca}{II} & 275 $\pm$ 15  & 30 $\pm$ 7  & 1.93201522\tablenotemark{a}\\
\Ion{Na}{I}  & 409 $\pm$ 26 & 34\tablenotemark{b}& 1.93201522\tablenotemark{a}
\enddata
\tablenotetext{a}{Given by the updated ephemeris.}
\tablenotetext{b}{Fixed to the $\gamma$ of H$_{\beta}$. }
\end{deluxetable}

\begin{figure*}
   \centering
   \includegraphics[width=16cm]{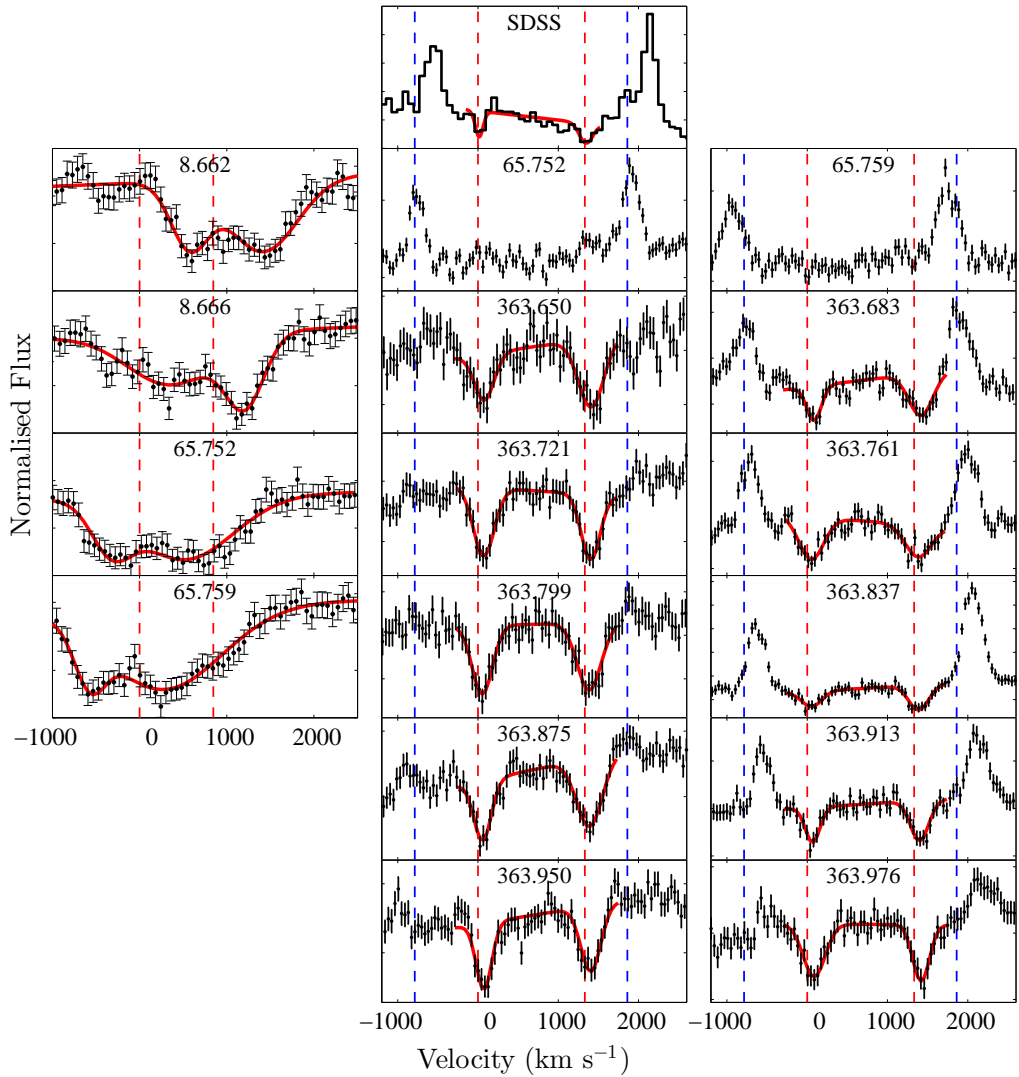}
   \caption{
     Left column: the region of the spectra around the
     \Lines{Na}{I}{8183.26,8194.81} absorption doublet for
     observations in which it was detected. Centre and right columns: the
     region of the spectra around the \Lines{Al}{I}{3944.01,3961.52} doublet, with
     the 2004 SDSS observation in the top-most panel.  In all cases,
     the solid red curves show our best fits to the features, and the
     red vertical dashed lines mark the zero-velocity positions of the
     doublets, and the relative heliocentric Julian date (HJD-2456000) is given in
     the top of each panel. The regions around the aluminum doublet
     also contain the \Lines{Ca}{II}{3933.66,3968.47} lines, which
     are sometimes seen in emission; the blue vertical dashed lines
     mark the expected zero-velocity positions of those calcium
     features.
   \label{NaAl}
    }
\end{figure*}

\begin{figure}
   \centering
   \includegraphics[width=0.45\textwidth]{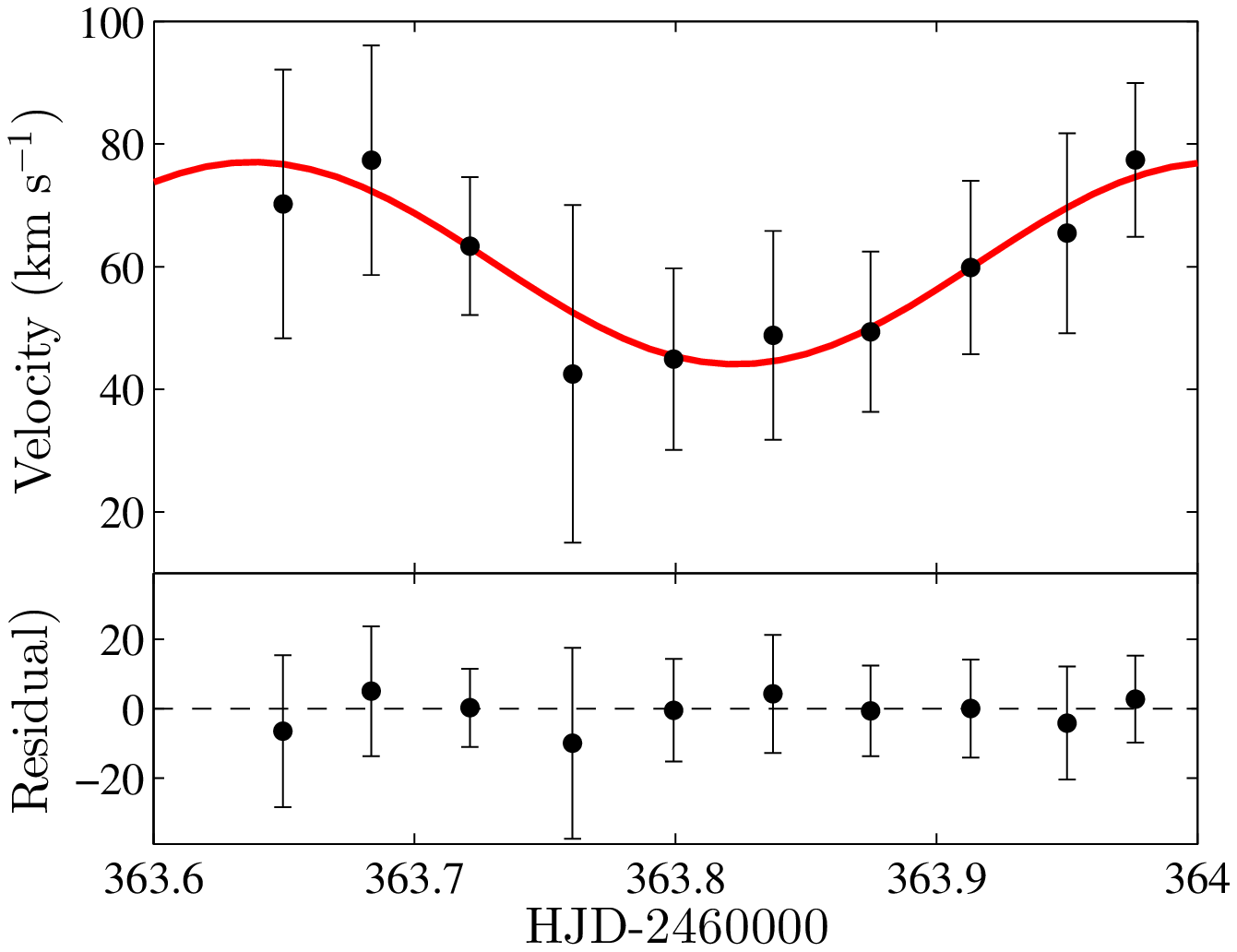}
   \caption{ \emph{Top}: the radial velocity of \Ion{Al}{I} doublet; the
     red curve shows our best fit to the data.
             \emph{Bottom}: the residuals from our fit.
    }
    \label{Al_rv}
\end{figure}

\section{Observations \& basic orbital parameters}
Optical spectra of AR UMa were obtained with the Double Spectrograph (DBSP) at the
Palomar Observatory. The dichroic D55, which splits light into blue and red channels around 5500 {\AA},
was used with two 1200 line mm$^{-1}$ gratings blazed at 5000 and 9400 {\AA}, respectively.
After each exposure of AR UMa, two exposures each of FeAr (blue channel) and
HeArNe (red channel) lamps were taken to minimize the systemic uncertainty
in the wavelength calibration.
The standard stars Hilter 600 and Hz 44 were observed before and after
the exposures of AR UMa. All observations are summarized in Table~1.

The spectra were reduced in a standard way using IRAF procedures.
After bias subtraction and flat correction, the one-dimensional (1D)
spectra were extracted with an aperture size of 10 pixels.
The wavelength calibration of each exposure was carried out using
the adjacent arc lamp exposures.
Those 1D spectra were then corrected for atmospheric extinction,
and flux-calibrated using the exposures of the standard stars.

\subsection{Updated Orbital Ephemeris}

The H$_{\beta}$ emission line is the strongest feature in all of our 16 observations.
After barycenter correction, we used a Gaussian function plus a parabola \citep{Rebassa07}
to fit the H$_{\beta}$ emission line and find its radial velocity (and error) in
each observation. These values are presented in Table~\ref{Pro}.
Assuming a circular orbit \citep{Howell01}, we then
fitted a sine curve to the radial velocities, keeping the period fixed
to 0.08050074(12) day \citep{Schmidt99} and iteratively minimizing $\chi^{2}$.
The best solution has a radial velocity amplitude of $K = 245 \pm 4$ \kms
and a systemic velocity of $\gamma = 36 \pm 5$ \kms (quoting 90\% uncertainties).

From this best-fit curve, we took five instances of positive zero-crossing of the radial velocity curve,
HJD (2456000 +) 363.641, 363.721, 363.802, 363.882, 363.963 with an
uncertainty of 0.008 days.
Combining this with the seven instances of positive zero-crossing in
\citet{Schmidt99}, we derive a best-fit orbital ephemeris of
\begin{equation}
  \mathrm{HJD} = 2450470.4314(4) + 0.080500634(8)\textit{E}.
\end{equation}

Fitting our H$_\beta$ emission line radial velocities again using this
updated orbital period finds $K = 248 \pm 4$ \kms and $\gamma = 34 \pm
5$ \kms (see Table~\ref{sin}).    We also fitted the \Ion{Ca}{II}
triplet with the same procedure as H$_\beta$, and find
a systemic velocity for the \Ion{Ca}{II} triplet that is
very similar to that of H$_\beta$ (with $\gamma = 30 \pm
7$ \kms), and a slightly higher radial velocity amplitude ($K = 275
\pm 15$ \kms).

\begin{figure}
   \centering
   \includegraphics[width=0.45\textwidth]{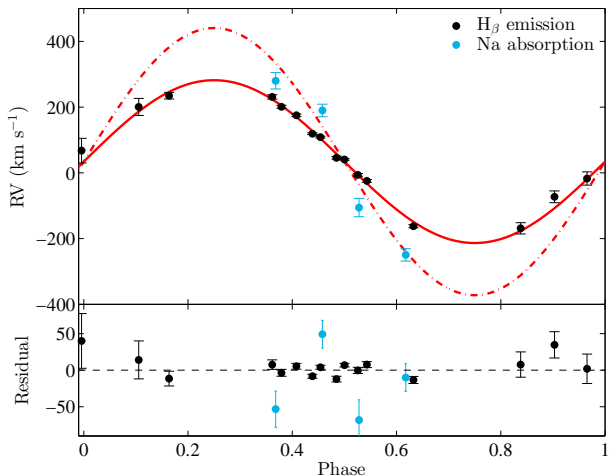}
   \caption{Top: the radial velocities for the H$_{\beta}$ and \Ion{Na}{I}
     lines versus the orbital phase, alongside our best-fit radial velocity
     curves (solid and dashed-dotted curves).
     Bottom: the residuals to our best-fitting models.
    \label{Phi1}
    }
\end{figure}

\section{Aluminum absorption doublet}
\label{sec:aluminium}

The \Lines{Al}{I}{3944.01,3961.52} doublet is present in the spectra from 2013 March.
We fit this doublet with a double-Gaussian function of a fixed separation
plus a parabola to measure the radial velocity and error of  the \Ion{Al}{I} doublet
(\citealt{Rebassa07}; see the middle and right panels in Figure~\ref{NaAl}).
The results of this fitting are presented in Table~\ref{Pro}.
We also attempted to fit the radial velocity curve of this \Ion{Al}{I}
doublet with a sine function fixed to the binary period of 1.932 hr,
but the results were unacceptable (with the $\chi^{2}$
roughly 10 times higher than that for the free-period fit).
The best-fitting parameters for an unconstrained period are listed in
Table~\ref{sin}, and the radial velocity curve is shown in
Figure~\ref{Al_rv}.  The total duration over which the relevant spectra were
taken (on 2012 March 3) was only $\approx$8 hr, so we do not at all claim that
this fitted period is definitive, or even that we have found
  evidence of any periodicity in the \Ion{Al}{I} doublet
  velocities.\footnote{However, we note that \citet{Kalomeni12}
  found evidence for a $\approx$7.9 hour period in AR UMa by
  examining light variations over 11 years. This is broadly similar
  to that of our best fit to the radial velocity variation in the \Ion{Al}{I} doublet,
  although we stress that our data so far are insufficient to claim
  detection of a true periodicity.}

The best-fit systemic velocity
($\gamma$) for the \Ion{Al}{I} doublet is clearly inconsistent with the
systemic velocity for the H$_{\beta}$ emission lines.  By itself, this
is not necessarily a strong constraint on the location of the
absorbing material within the system, since $\gamma$-velocities for
polars are known to suffer from complications which are not seen in the
population of non-magnetic CVs \citep[][]{vanParadijs+1996}.
Nonetheless, this result is consistent with our later conclusion that
the absorbing material is probably circumstellar.

We are confident that this \Ion{Al}{I} doublet is unlikely of telluric origin or an
artifact of the telescope system, nor is it from the foreground interstellar medium,
since the dispersion is about 300 \kms.
Moreover, we suggest it is not from the MD, since the radial velocity amplitude and
period are inconsistent with those of the \Ion{Na}{I} absorption line
from the MD (see the following section).

That \Lines{Al}{I}{3944.01,3961.52} doublet is not detected in the
spectra from 2012 May.  Unfortunately, there is no coverage
of this region in our spectra on HJD 2456008. So any inference of
qualitative time variability -- disappearance and reappearance -- rests
on upper limits derived for only two spectra from HJD 2456065 (see Figure~\ref{NaAl} and
Table~\ref{Pro}); one of those shows very weak hints of a doublet by eye,
but our fitting indicates an upper limit that is well below the later
detections. The simplest interpretation is that the column density
  of aluminum in absorption changed between our 2012 and 2013 observations.

An alternative option to consider is that during the time of the 
2012 May observations we are just not seeing the aluminum absorption.
However, these
spectra were taken at orbital phases of 0.54 and 0.63, and the
spectrum taken at HJD of 363.683 was at an orbital phase of 0.53, which
suggests that the disappearance of the aluminum absorption is not a
simple function of orbital phase. The 10
spectra from HJD 363 are spread across all of the orbit, and the Al
absorption lines are clearly detected in all of them.

We also find that these \Ion{Al}{I} absorption lines were detected in an
SDSS spectrum of AR UMa from 2004 February (Julian day 2453061.4). If
our observations are robust then any model for these lines needs to
explain the disappearance of aluminum features over less than 8.2
years, and their reappearance in less than a year, with no obvious
dependence on orbital phase. 

We have not detected other metal absorption lines from our spectra.
However, if they are present, some might perhaps have been disguised by
the emission lines from the MD (e.g., \Ion{Mg}{I} and \Ion{Ca}{II}).
Here we re-examine the far-ultraviolet spectra in \citealt{Gansicke01},
and fit the strongest metal absorption line of \Ion{Si}{II} with a Gaussian
function plus a line. The result shows that the FWHM is 240 $\pm$ 80 \kms, which
is similar to that of the aluminum absorption doublet.

\begin{figure*}
   \centering
   \includegraphics[width=0.8\textwidth]{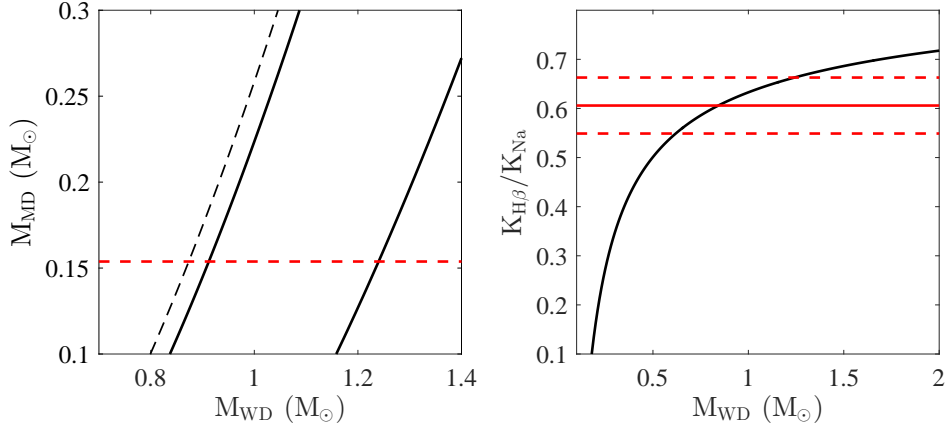}
   \caption{ Left panel: the mass of the MD as a function of that
     of the WD.  A given inclination angle fixes the mass ratio; the
     curves from left to right give the resulting loci in this plane for different
     inclination angles of 75\arcdeg, 71\arcdeg, and 57\arcdeg.
       The dashed horizontal line indicates the mass of the MD as found from the semi-empirical
        relation for CVs \citep{Knigge11}.
             Right panel: the inferred mass of the MD as a function of the amplitude ratio.
             The black solid curve represents the expected ratio between the
             radial velocity amplitudes from the L1 point and the
             barycenter of the MD ($K_{\mathrm{L1}}$/$K_{\mathrm{MD}}$).
             The ratio between the inferred H$_{\beta}$ and
             \Ion{Na}{I} radial velocity amplitudes
             is indicated by the red solid horizontal line (with the
             dashed red lines showing uncertainty ranges
             conservatively including $3
             \sigma$ uncertainty in both radial velocity amplitudes).
   \label{mr}
    }
\end{figure*}

\begin{figure}
   \centering
   \includegraphics[width=0.45\textwidth]{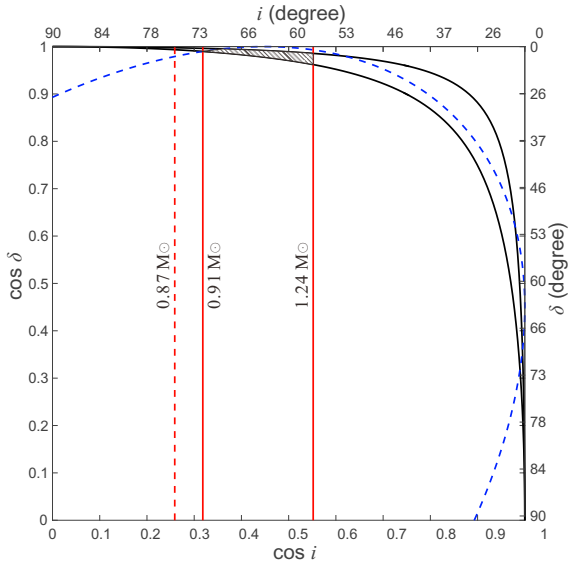}
   \caption{Constraints on the geometry of AR UMa. Only the
       small shaded region is allowed when combining multiple published constraints.
            The dashed red vertical line marks $i$ = 75\arcdeg, and the solid red vertical lines mark
            $i$ = 71\arcdeg and 57\arcdeg.
            The black solid and blue dashed curves represent
            constraints inferred from the polarization data, as
            described in \citealt{Schmidt99}.
   \label{geo}
    }
\end{figure}

\begin{figure*}
   \centering
   \includegraphics[width=0.8\textwidth]{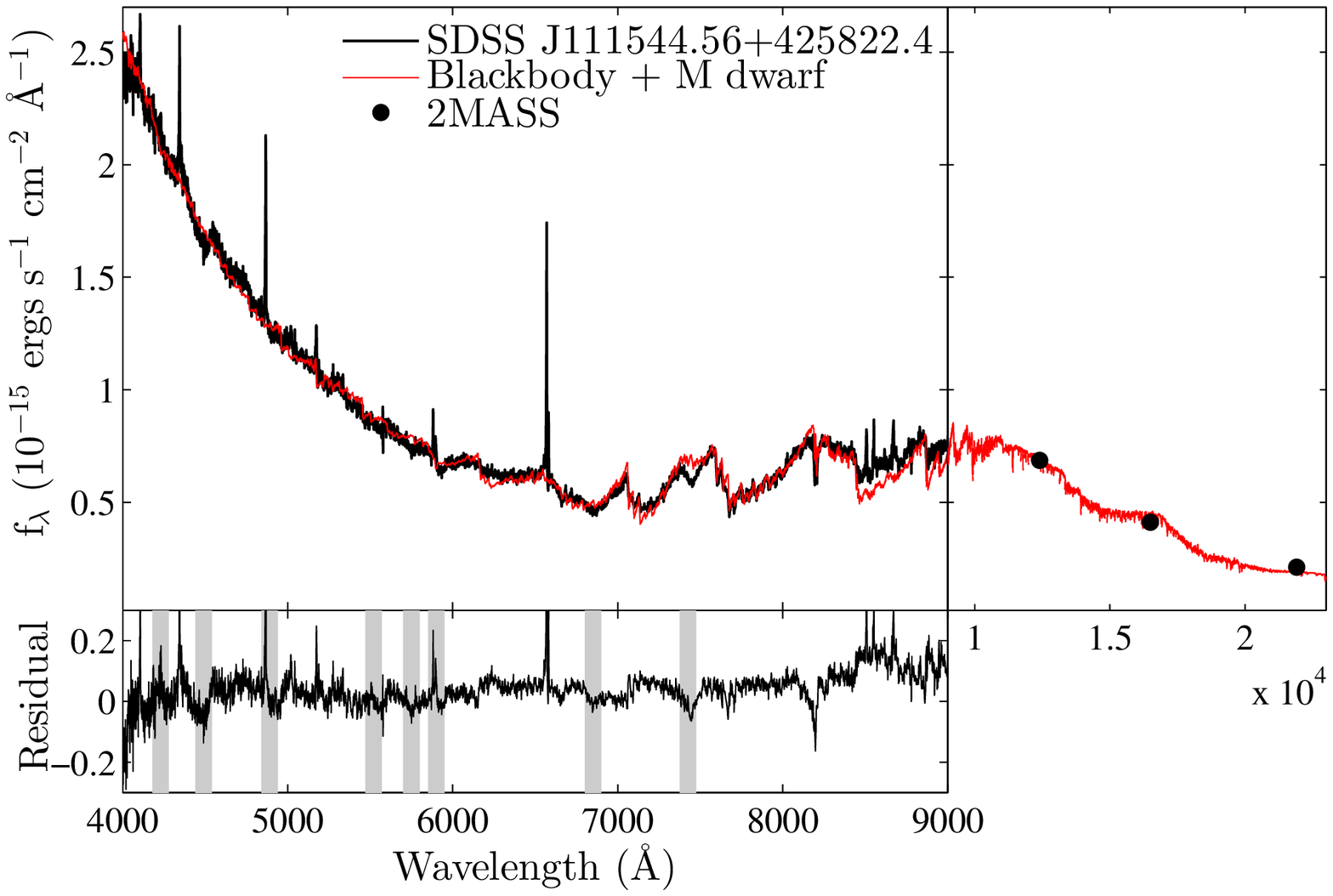}
   \caption{ Top: our best-fit model spectrum (red line)
     compared to SDSS observations of AR UMa (black line) and 2MASS
     photometry (black circles).
             Bottom: the residual to the fits (black line).
             Regions in which polarization is expected to affect the
             spectral features are indicated by the gray shaded zones.
   \label{sdssfit}
    }
\end{figure*}

\section{Stellar Parameters}

AR UMa was in a low brightness state during our observations, since
the molecular band from the MD was obviously presented and the continuum flux
was similar to that during the low brightness state observed by \citet{Schmidt96}.
It was also probably in a low accretion state, since the \Ion{He}{II} emission
line was detected in only one observation, and only with a signal-to-noise ratio below 3,
which suggests that the accretion rate is minimal \citep{Szkody99}.
In the low state, the emission lines are thought to originate from the inner hemisphere of the
MD, rather than the directed stream flow \citep{Schmidt99}.

These low-state observations enable us to trace the radial velocity of the MD, and
further constrain the mass of the WD with the mass and the radial velocity amplitude
of the MD.

\subsection{Mass of the WD}

The emission lines of AR UMa originate from the hemisphere of the MD facing the
WD, so their velocities' amplitudes should be smaller than that of the MD barycenter
\citep{Schmidt99}.
In AR UMa, the photospheric \Lines{Na}{I}{8183.26,8194.81} doublet is more
isotropically distributed over the MD \citep{Rebassa07,Schwope11}.
This is detected in spectra from the red channel in four of the 2012 observations.
We fit the doublet with a double-Gaussian function of a fixed separation
plus a parabola to measure the radial velocity and error (the left panel in Figure~\ref{NaAl}).
We then fit the velocity curve of \Ion{Na}{I} with a sine function,
keeping $\gamma$ and $P$ fixed to the values derived
from the radial velocity of H$_{\beta}$ and the updated orbital ephemeris.
The radial velocity amplitude derived from the best fitting is $K =$ 409 $\pm$ 26 \kms
(Figure~\ref{Phi1}), i.e., higher than for the H$_{\beta}$ lines, as expected.

We thereby estimate the mass of the MD to 0.154 M$_{\odot}$ from the
semi-empirical relation for CVs \citep[see][]{Knigge11}. This
semi-empirical relation is typically taken to have an expected
uncertainty below 2\% \citep{Knigge11}, although the extreme
nature of AR UMa might plausibly make this estimate less
reliable than for normal, non-magnetic CVs.

Since no eclipse is seen from AR UMa, the inclination angle $i$ should
be smaller than 75\arcdeg \citep{Schmidt99}, which constrains the mass of the WD
to $M_{\mathrm{WD}}$ $>$ 0.87 $M_{\odot}$.
\citet{Harrison15} found a best-fit inclination for AR UMa of 65\arcdeg,
which would correspond to a WD mass of 1.01$_{-0.14}^{+0.16}$ M$_{\odot}$.
The uncertainty adopted here is calculated from the 90\% uncertainty of the
\Ion{Na}{I} amplitude. This should not be treated as giving real upper and lower
limits, since no uncertainty is included for the model-fit inclination
angle.

Since we expect that the barycenter of the irradiated H$_{\beta}$ emission area
should be located between the inner Lagrange point (L1) of the binary
and the barycenter of the MD, this may help to constrain
the mass of the WD.
We first compute the ratio of the radial velocity amplitudes for the L1
point and the MD barycenter (i.e.,
$K_{\mathrm{L1}}$/$K_{\mathrm{MD}}$) as a function of the mass of the
WD (assuming the MD mass found above), as plotted in the right panel
of Figure~\ref{mr}. Our fits to the observed radial velocity data lead to an amplitude
ratio ($K_{\mathrm{H\beta}}$/$K_{\mathrm{Na}}$) of
0.606 $\pm$ 0.057, where the quoted uncertainty range conservatively adopts $3
\sigma$ uncertainties in both the radial velocity amplitudes.
The upper end of that range of statistical uncertainty corresponds to an upper
limit on the WD mass of $M_{\mathrm{WD}}$ $<$ 1.24 $M_{\odot}$, corresponding to
an inclination angle of 57\arcdeg.

This excludes a possible source of systematic uncertainty,
since the \Lines{Na}{I}{8183.26,8194.81} doublet may trace the outer
hemisphere of the MD, i.e., be biased toward the
side away from the L1 point. Nonetheless we still
feel that this is probably a fairly conservative limit, as it assumes
that the H$_{\beta}$ lines trace the L1 point.

Combining this limit with the lower limit from taking
$i$ $<$ 75\arcdeg~produces an estimate for the
mass range of the WD in AR UMa of 0.87 $M_{\odot}$ $<$
$M_{\mathrm{WD}}$ $<$ 1.24 $M_{\odot}$.

\subsection{System Geometry}
\citealt{Schmidt99} estimated the possible ranges of the
  inclination angle $i$ and the colatitude
of the magnetic pole $\delta$ using constraints calculated from photospheric
circular polarization. Combining their results with ours
($K =$ 409 $\pm$ 26 \kms, 0.87 $M_{\odot}$ $<$ $M_{\mathrm{WD}}$ $<$ 1.24 $M_{\odot}$),
we update their constraints on $i$ and  $\delta$ to
\begin{equation}
  57\arcdeg < i < 71\arcdeg, 10\arcdeg < \delta < 16\arcdeg,
\end{equation}
In turn, this lower upper limit on $i$ slightly increases the minimum inferred
mass of the WD; the allowed region in Figure~\ref{geo} yields a WD
mass range of 0.91 $M_{\odot}$ $<$ $M_{\mathrm{WD}}$ $<$ 1.24 $M_{\odot}$. 

Such system geometry (67\arcdeg $<$ $i +$ $\delta$ $<$
  87\arcdeg) would imply that
one hotspot around the magnetic pole of AR UMa is always visible.

\subsection{Spectral Decomposition}

We also constructed two-component (blue and red)
models based on WD and MD templates to fit our observed spectra.
Since the Balmer absorption lines of the WD were undetected in the blue
component of AR UMa, a series of blackbody spectra were used as WD
templates. We also note that \citet{Schmidt96} previously suggested
that the blue component contained contributions from both the inner hemisphere of the
MD and the atmosphere of the WD.
For the MD templates, we adopt the Phoenix library of synthetic spectra \citep{Husser13}.
The model space therefore covers a WD parameter {\Tbb} and MD parameters {\Teff},
log $g$, and [Fe/H]. The relevant ranges in the parameter space are:
WD blackbody temperature 10 kK $\leq$ \Tbb $\leq$ 50 kK
(step size 0.1 kK); effective temperature 2300 K $\leq$ \Teff $\leq$ 3400 K
(step size 100 K); surface gravity 0.0 $\leq$ log $g$ $\leq$ 6.0
(every 0.5 dex); metallicity $-$4.0 $\leq$ [Fe/H] $\leq$ 1.0 (every 0.5 dex).

Since our spectra have limited coverage, we used the SDSS spectrum of AR UMa
(SDSS J111544.56+425822.4) and minimized the $\chi^{2}$ parameter to
fit the spectrum in a four-dimensional parameter space, after masking
all of the emission lines in the spectrum.
The spectrum is reproduced at each point in the parameter space by a combination of
a blackbody and a MD. They are then weighted with scaling factors that depend on the
distance and the radii.
The fitted spectrum (Figure~\ref{sdssfit}) gives {\Tbb} $=$ 46300 K, {\Teff} $=$ 3200 K, log $g =$ 5.0 and
[Fe/H] = 1.0. The half widths of the parameters are adopted as the uncertainties \citep{Liu12},
which are 50 K for {\Tbb} and {\Teff}, 0.3 dex for log $g$ and 0.3 dex for [Fe/H].

The inferred temperature and gravity of the MD are consistent
with those calculated from the semi-empirical relations of CVs
({\Teff} $\simeq$ 3180 K and log $g$ $\simeq$ 5.1).

However, the fitted  blackbody temperature is much higher than previously estimated for the
WD in AR UMa;  for example, \citealt{Gansicke01}, estimated a WD temperature
of $20,000 \pm 5000$ when including information from ultraviolet HST spectra of AR
UMa.

The scaling factors enable us to obtain the radius of the MD and the
effective radius of this blackbody component (assuming a spherical emission
surface). To do this we adopt the distance of AR UMa, 86 (+10, $-$8) pc, measured from its
parallax \citep{Thorstensen08}.
The derived radii are 0.15 $\pm$ 0.02 R$_{\odot}$ for the MD,
and 0.0034 $\pm$ 0.0004 R$_{\odot}$ for the blue component.
The radius of a non-magnetic WD with 1.01 M$_{\odot}$ is in the range
of 0.0074 $<$ R$_{\mathrm{WD}}$ $<$ 0.0084 R$_{\odot}$,
taking effective temperatures between 10000 K and 40000 K from
non-magnetic DA WD models
\citep{Holberg06,Tremblay11}\footnote{http://www.astro.umontreal.ca/$\sim$bergeron/CoolingModels/}.
Assuming a strong internal magnetic field for the WD in AR UMa would
increase the expected radius \citep{Suh00}, so the emission area of
the fitted blue component is too small to cover the complete surface of the WD
(unless the WD is substantially more massive than we have inferred).

The inferred radius of the MD is smaller than that estimated using the semi-empirical
relation for CVs, $R$ $\simeq$ 0.19 R$_{\odot}$.  This underestimate seems
consistent with the red component of the continuum being dominated by
the outer hemisphere of the MD.

In Figure~\ref{sdssfit}, the residuals from the fitting show features
that clearly deviate
from the synthetic model. Some of these are due to the polarizing
effect of the strong magnetic field \citep{Schmidt96,Ferrario03}.
The deviation around 8200{\AA} is due to the variability of the \Ion{Na}{I}
line profile (Figure~\ref{NaAl}), and
the continuum around \Ion{Ca}{II} triplet may be due to
the unresolved Paschen emission lines.

\section{Discussion}
We have detected an aluminum absorption feature in spectra of AR
UMa, during the probable low brightness and accretion states. The
apparent sub-year time variability of the feature is especially intriguing.
Here we discuss potential origins of the aluminum absorption.

\subsection{Interpreting the \Ion{Al}{I} doublet as from circumstellar material}

Since the radial velocity of the \Ion{Al}{I} doublet is inconsistent with that of the MD,
the aluminum is unlikely to be located on the surface of the MD.
\citealt{Gansicke01} detected a P Cygni profile for Ly$\alpha$ emission, 
suggesting the existence of a moderately fast wind, about $-$700 \kms, from the MD,
which is also inconsistent with the velocity of the \Ion{Al}{I} doublet.  

Another argument against the aluminum being located on the
  donor star is the sub-year timescale over which we see the feature
  changing, given that the variation does not seem to be due to orbital
  changes (see Section \ref{sec:aluminium}). This is not a natural
  timescale for variations in the composition of the MD atmosphere.

The $\lessapprox8$ year timescale for disappearance of the aluminum
  feature
  could be easily explained by gravitational settling in the
  atmosphere of the WD. The reappearance of the aluminum in less than a
  year might also be naturally explained by pollution of the WD.
Recent years have seen great interest
in studying WDs with atmospheric heavy elements, especially because
these heavy elements may have been delivered by the accretion of rocky
material from a disrupted planet, debris disk, or asteroids
\citep{Zuckerman07,Zuckerman11,Dufour10,Dufour12,Pyrzas12}.
Previous observations have indicated time variation in such
accretion onto some isolated WDs, perhaps due to a transient gaseous
disk \citep{Wilson14} or precession of the debris disk
\citep{Manser15}.

An alternative explanation for the presence of
atmospheric heavy elements in suitably hot WDs is radiative levitation
\citep[see, e.g.,][]{Chayer+1995}.
If the observed variability had been due to the changing effect of radiative
levitation on the WD, this might have suggested that the
effect was rather sensitively balanced on the WD in AR UMa.

However, the profile of the \Ion{Al}{I} doublet strongly argues against
  the absorption feature being from the surface of the highly magnetic
  WD in AR UMa. All the spectral features from the atmosphere of the
  WD should be split into wide ranges
of wavelength and be barely recognizable \citep{Kepler13} for AR
UMa. Assuming a dipole
magnetic field with a strength of $\sim$ 240 MG at the poles gives a
field strength of $\sim$80$-$120
MG at the equator \citep{Martin84,Gansicke01}.  We consider it highly
unlikely that a sufficiently large patch of the WD surface would have
low enough magnetic field strength for the \Ion{Al}{I} doublet to be
so clearly seen with no Zeeman splitting.
We also note that our attempts to fit the \Ion{Al}{I} absorption line with a
radial velocity curve did not produce anything consistent with that expected for the WD.
Overall, we consider the lack of magnetic distortion in the lines from
  the \Ion{Al}{I} doublet to be convincing evidence that the lines are
  not formed on the surface of the WD.

Given the above, the two remaining possibilities are
  interstellar or circumstellar absorption. Based on the observed
  variability, we suggest that the absorption is unlikely to be
  interstellar. So we conclude that the absorption features probably
  have a circumstellar origin.

\citet{Gansicke01} found weak low-ionization lines in their FUV spectrum
of AR UMa, which they interpreted as likely being due to the absorption of
the local interstellar medium.
Perhaps the \Ion{Al}{I} absorption doublet is from similar material as
provided the origin for those low-ionization lines, in which case
\citet{Gansicke01} also detected such circumstellar gas.

\subsection{A rocky origin for the circumstellar material?}

\begin{figure}
   \centering
   \includegraphics[width=0.4\textwidth]{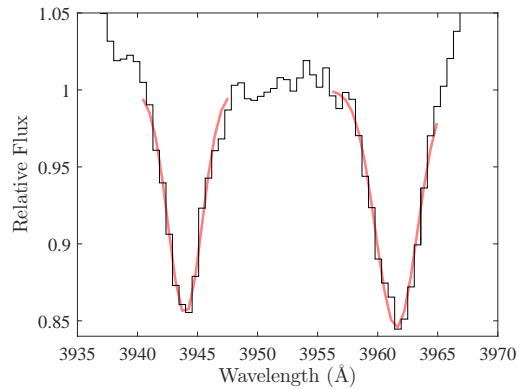}
   \caption{ The fit (red line) of \Ion{Al}{I} doublet for the
     center-corrected combined spectrum (black line) with \texttt{VPFIT}.
   \label{eva}
    }
\end{figure}

Assuming we have detected circumstellar gas in AR UMa, we now
discuss the likely origin of that material. Circumstellar gas around WDs
has previously been reported in four systems: WD 1040$+$492, WD 1942$+$499,
WD 2257$+$073 and WD 1124$-$293 \citep{Lallement11,Debes12}.
An anti-correlation is derived between the effective temperature of the
WD and the orbital velocity of the gas around the WDs, which is consistent
with the orbiting gas material originating from the evaporation of planets or
asteroids.

Additional evidence for such a scenario follows from several
  transit events of planets discovered from the light curve of WD
  1145$+$017 \citep{Vanderburg15}. The varying transit depths and
  their asymmetric profiles indicate a planet with a comet-like tail
  from evaporation.  The tail, consisting of various heavy elements,
  would escape into the orbit through a Parker-type thermal wind \citep{Rappaport14,Sanchis15}.
  In that case the tail also contaminates the atmosphere of WD 1145$+$017.

Besides isolated WD systems, three polars -- HU Aquarii, UZ Fornacis,
and CSS081231:071126+440405 -- are reported as having circumbinary planets
\citep{Potter11,Qian11,Schwope15}, potentially in stable highly non-coplanar orbits
\citep{Gozdziewski15}.

If the inferred circumstellar gas in AR UMa did originate from the
  evaporation of a rocky body, measuring the mass-loss flux might help to constrain the
  characteristics of the body evaporation \citep[see especially][]{Rappaport+12,Perez13}.
   
We can estimate the column density of
\Ion{Al}{I} in 2013 May 12 by taking the center-corrected spectra from
those observations and fitting the doublet with \texttt{VPFIT}
(Figure~\ref{eva}, \citealt{Carswell14}).
The best-fitting column density is $N$(\Ion{Al}{I}) = 3.9 $\pm$ 0.2
$\times$ 10$^{13}$ cm$^{-2}$. Unfortunately, the spacing of our
observations does not sufficiently
constrain the timescale of evaporation to provide a useful estimate of
the mass-loss flux.



We estimate that the region in which an aluminum-rich rock would be
  expected to sublimate is much larger than the volume in which Zeeman
  splitting would have been detected in our low-resolution
  spectra (corresponding to a field strength of roughly 2MG, and a
  distance from the WD of $\approx 0.05 R_{\odot}$). Taking a
  sublimation temperature for Corundum (Al$_2$O$_3$) of 1500K, we find
  that the equilibrium temperature caused by heating from the WD would
  exceed this value within $\approx 1.6 R_{\odot}$ from the WD. This is
  very similar to the major axis of the orbit of the MD.   Moreover,
  this is comparable to the separation at which the WD would tidally
  disrupt such an asteroid (assuming a density
  of  4.02 g cm$^{-3}$ for Corundum); for asteroids with lower
  density, tidal disruption would likely begin before evaporation
  \citep[see, e.g.,][]{Veras14}.

We note that the relative radial velocity of the fitted
  \Ion{Al}{I} doublet to the AR UMa binary system
is $\approx$ 27 \kms, which is consistent with circumstellar gas
detected around other WDs \citep{Lallement11}.



We do not claim that it is inevitable that this interpretation
  of circumstellar aluminum demands a rocky origin.  An alternative
  possibility might be related to a proposed explanation for the known
  CVs with suprasolar photospheric abundances of aluminum
  \citep{Sion01,Gansicke05,Jose06}. Some models for those CVs
  indicate that aluminum was likely generated in
  nova outbursts on the WD, and furthermore that the
  aluminum-enriched nova ejecta might have
  been captured by the donor star and re-accreted onto to the WD
  \citep{Marks98,Sengupta13,Sion14}.  If our interpretation
that the aluminum in AR UMa is circumstellar is correct, then it might perhaps
be explained by a variation of that model, if some of the nova ejecta
can remain within the system.  However, it is less
obvious to us how the time variation we observe might be naturally
explained in this model.


So a tentative potential explanation for the origin of this circumstellar gas is that
AR UMa possesses an extrasolar analog to the Oort cloud \citep{Stone15} or the
Kuiper Belt \citep{Bonsor11} containing some Al-rich rocks.
The orbits of those rocks might become destabilized due to perturbation of the binary motion or
dynamical interaction with other planets in the system \citep{Debes12b,Frewen14},
after which they approach so close to the WD as to be evaporated, resulting in circumstellar gas
in the form of a highly eccentric ring \citep{Veras14,Veras15} or spherical cloud
around the WD \citep{Stone15}.
That circumstellar gas may accrete on to the WD with a time-scale of less than $\sim$ 8.2 years
due to the additional force from the MD \citep{Veras14},
which might explain the time-scale over which the aluminum absorption
disappears.

If this scenario is correct, more time-resolved spectra of these lines
may provide a way to probe the duty cycle of the evaporation and further shed light
on the population of asteroids in AR UMa and their orbital instabilities.
Additional infrared spectroscopy might constrain the
composition of the gas by detecting molecular features and thus investigate
the character of these asteroids \citep{Carry12}.

\section{Summary}

We have detected an aluminum absorption doublet in spectra of AR UMa,
and suggest that this is evidence for circumstellar material
  in this extreme polar.
The aluminum
absorption feature also appears to be time-variable, having disappeared
over less than 9 years, and then re-appeared in less than a year.  Our
spectra also allowed a preliminary attempt to better-constrain the
properties of the components in AR UMa, including an updated estimate
for the WD mass in the range of 0.91 $M_{\odot}$ $<$ $M_{\mathrm{WD}}$ $<$ 1.24 $M_{\odot}$.

A potential explanation for this variable aluminum absorption is irregular
circumstellar gas due to evaporation of rocky material from, e.g.,
asteroids destabilized from orbits further out in the system.
We encourage studies of AR UMa with high time-resolution spectra, both to enable
deeper investigation of the origin of this absorption and to help
further improve our understanding of this extreme system.

\begin{acknowledgements}
We sincerely thank Patrick Dufour for valuable discussions, especially for
  important clarifications regarding the Zeeman effect in the very strong
  field regime. We are also grateful to an anonymous referee for
  forcing us to re-examine our initial interpretation, and to Koji
  Mukai, Jay Farihi, and Rosanne Di Stefano for very helpful thoughts.
This research uses data obtained through the Telescope Access Program (TAP),
which is funded by the National Astronomical Observatories and the Special Fund
for Astronomy from the Ministry of Finance.
This work was supported by the Chinese National Natural Science Foundation (NSFC)
through grants NSFC-11333004/11425313, and the National Astronomical Observatories,
Chinese Academy of Sciences under the Young Researcher Grant. S.J.
thanks the Chinese Academy of Science for support through the President's
International Fellowship Initiative grant No.~2011Y2JB07.
\end{acknowledgements}

~~~~~~~~~~
\vspace{30ex}

\begin{figure*}
   \centering
   \includegraphics[width=0.9\textwidth]{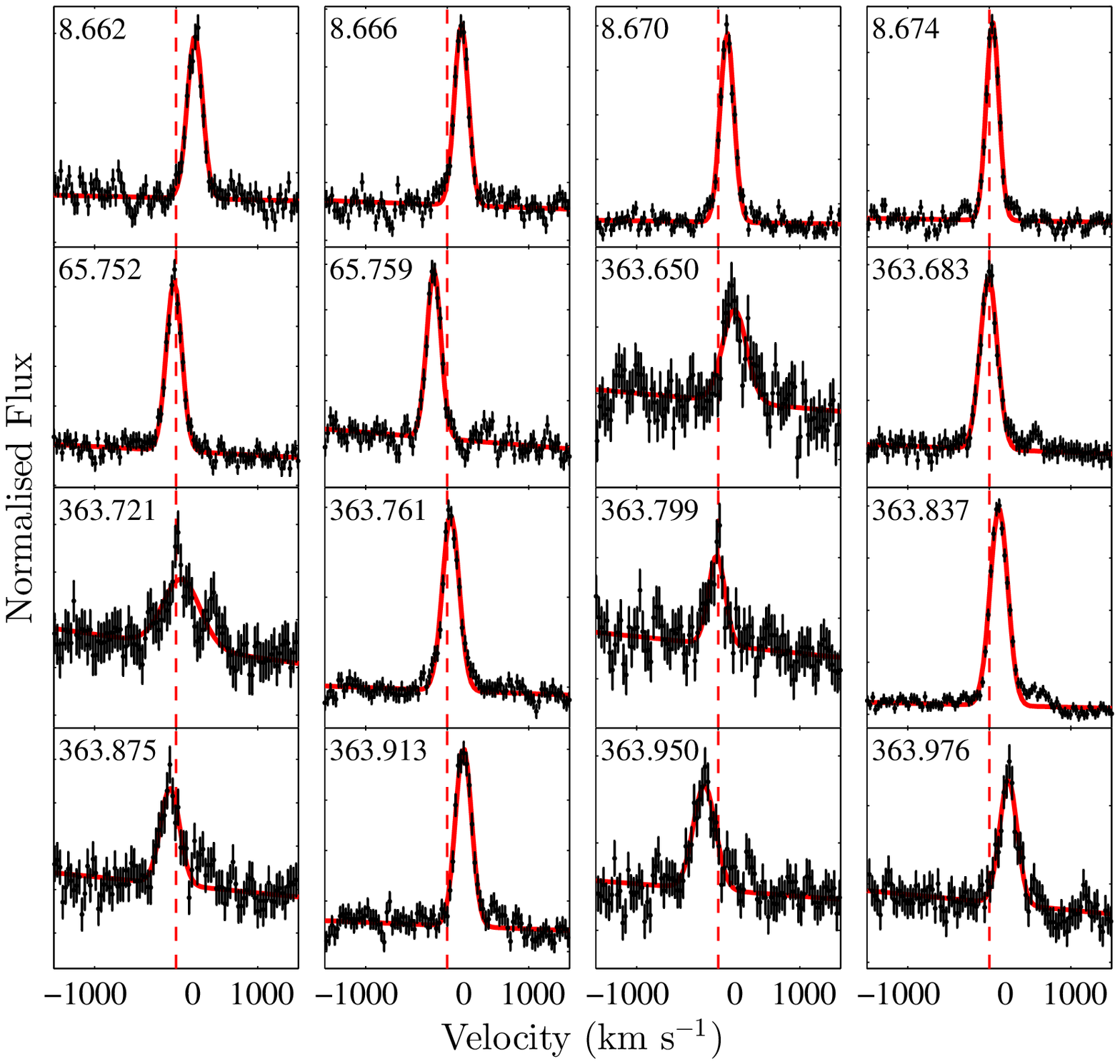}
   \caption{ Region of the spectra around the H$_\beta$ emission line.
    The best fits are shown as red solid lines, and the red vertical dashed lines
    mark the zero-velocity positions. The heliocentric Julian date-2456000 is given
    in the top left corner of each panel. }
    \label{Hbeta}
\end{figure*}

\begin{figure*}
   \centering
   \includegraphics[width=0.9\textwidth]{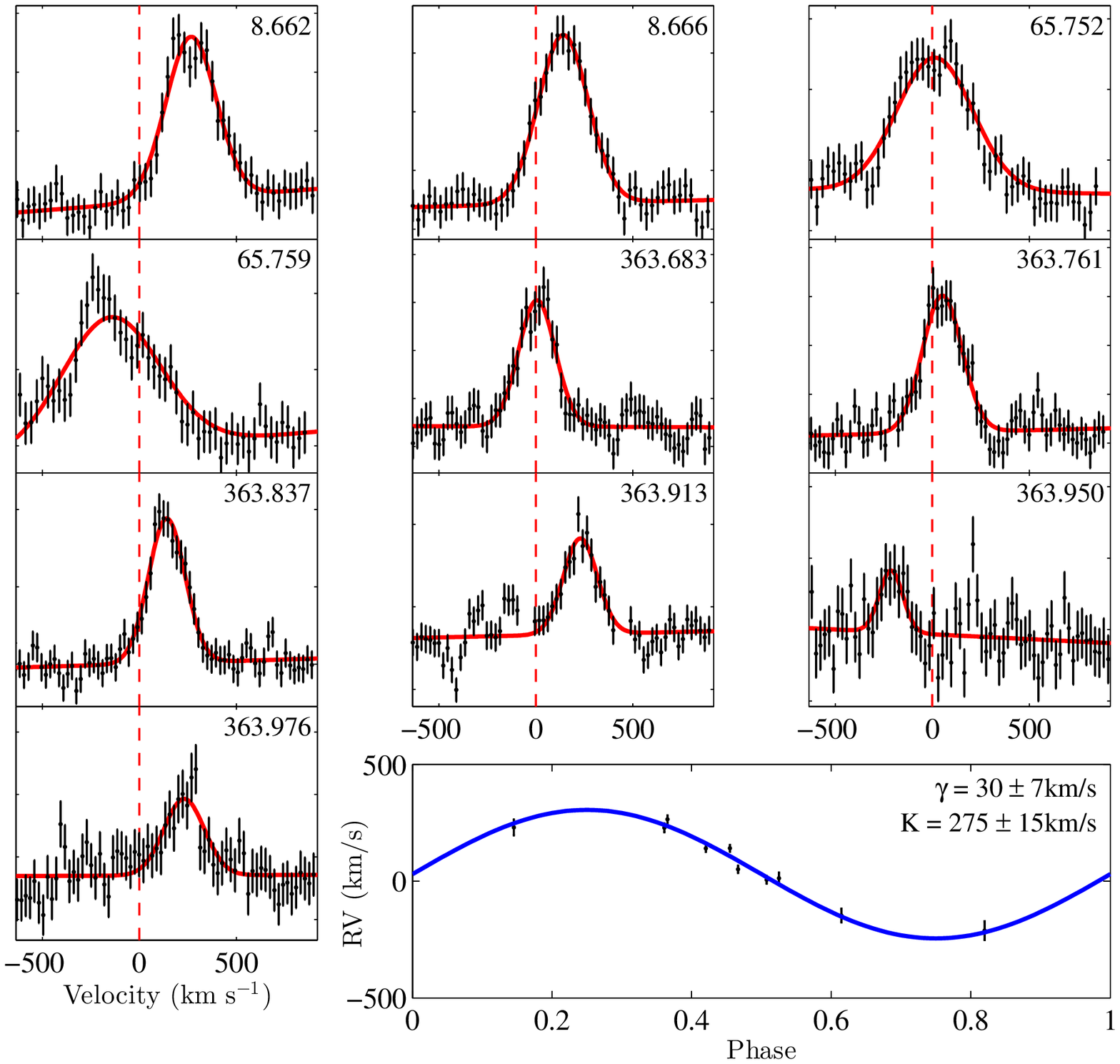}
   \caption{ Region of the spectra around $\lambda$ = 8498 \AA.
    The best fits are shown as red solid lines for the first line of
    the \Ion{Ca}{II} triplet, and the red vertical dashed lines mark the
    zero-velocity positions. The heliocentric Julian dates-2456000 are given
    in the top right corners of the first 10 panels. Our best-fitting
    radial velocity (RV) curve is shown as the blue solid line in the bottom right panel. The fitted
    systemic velocity ($\gamma$) and the radial velocity amplitude ($K$) are presented
    in the top right corner of the panel.}
    \label{Ca3}
\end{figure*}

\section*{Appendix}
We present the fits of the H$_\beta$ emission lines in our data (see Fig.~\ref{Hbeta}), and of
the first line ($\lambda$ = 8498 \AA) in the \Ion{Ca}{II} triplet (see
Fig.~\ref{Ca3}). For a radial-velocity fit to the \Ion{Ca}{II} triplet
in Fig.~\ref{Ca3}, we keep $P$ fixed to the value from the updated orbital ephemeris.

\end{document}